\documentclass[12pt]{article}

\usepackage{pb-diagram}
\def\oplusinf{\mathop{\oplus}}
\def\im{{\mbox{Im}}}
\def\ker{{\mbox{Ker}}}
\usepackage{amsmath}
\usepackage{amssymb}
\usepackage{amscd}
\usepackage{float}
\usepackage{theorem}
\usepackage{graphics}
\usepackage{graphicx}

\newtheorem{theorem}{Theorem}
\newtheorem{lemma}{Lemma}

\begin{document}

\begin{titlepage}

\begin{centering}

{\huge {\bf $N$-Complexes and Higher Spin Gauge Fields\footnote{Presented at the International Workshop ``Differential Geometry, Noncommutative Geometry, Homology and Fundamental Interactions" in honour of Michel Dubois-Violette, Orsay, April 8-10, 2008.}
}}

\vspace{2cm}

{\Large Marc Henneaux} \\

\vspace{1cm}

Physique th\'eorique et math\'ematique, Universit\'e Libre de Bruxelles $\&$ \\ International Solvay Institutes,\\ ULB-Campus Plaine CP 231, B-1050 Bruxelles, Belgium
\\

\vspace{.2cm}
Centro de Estudios Cient\'{\i}ficos, Casilla 1469, Valdivia, Chile 
\vspace{1.5cm}

\end{centering}

\begin{abstract}
$N$-complexes have been argued recently to be algebraic structures relevant to the description of higher spin gauge fields.  $N$-complexes involve a linear operator $d$ that fulfills $d^N = 0$ and that defines a generalized cohomology. Some elementary properties of $N$-complexes and the evidence for their relevance to the description of higher spin gauge fields are briefly reviewed.
\end{abstract}

\vfill
\end{titlepage}

\section{Introduction}	

In spite of crucial insights and advances in the development of higher spin gauge fields by the Vasiliev school (for reviews, see \cite{Vasiliev:2004qz,Vasiliev:2004cp,Bekaert:2005vh}), it is fair to say that some work remains to be done in order to reach a deeper and streamlined understanding of the algebraic structures underlying these physical systems (see \cite{Bekaert:2008sa}).  An important motivation for getting a better control of higher spin gauge fields comes from the study of hidden symmetries of gravity and hyperbolic Kac-Moody algebras where these fields seem to play an important role \cite{Julia,Julia2,West,DamourHN,HenryLabordere:2002dk,Henneaux:2007ej}.

A few years ago, it has been argued that $N$-complexes and their generalized cohomology are important ingredients in the description of higher spin gauge fields \cite{DuboisViolette:1999rd,DuboisViolette:2001jk}.  A $N$-complex is a graded vector space equipped with a $N$-nilpotent linear operator $d$, i.e., an operator that fulfills
\begin{equation}
d^N = 0 \, .
\label{dN=0}
\end{equation}
Given a $N$-complex, one can define generalized cohomologies
\begin{equation}
H_{(k)} \equiv \frac{\ker \, d^k}{\im \, d^{N-k}}
\label{coho}
\end{equation}
($k = 1, \cdots, N-1$) and establish useful relations among them \cite{DV1,DV2,DV3}.

The purpose of this short review is to briefly recall how the spaces of higher spin gauge fields are naturally equipped with such structures and to survey the relevance of the corresponding cohomologies.  This review is based on the papers \cite{DuboisViolette:1999rd,DuboisViolette:2001jk} jointly written with Michel Dubois-Violette, pioneer in the field, to whom it is a pleasure to dedicate this article.  Generalizations of the structures considered here have been discussed in \cite{Olver,Bekaert:2002dt}.

\section{Higher Spin Gauge Fields}

Classical spin $S$ gauge fields (with $S \in  \mathbb{N}$) are described by symmetric tensor fields $h_{\alpha_1\cdots \alpha_S}$ of order $S$ and gauge transformations of the form
\begin{equation}
\delta_\epsilon h_{\alpha_1\cdots \alpha_S} = \partial_{(\alpha_1}\epsilon_{\alpha_2 \cdots \alpha_S)} \label{gauge}
\end{equation}
where $\epsilon_{\alpha_2 \cdots \alpha_S}$ is a symmetric tensor of order $S-1$ (Young tableaux with $S$ or $S-1$ columns and 1 row, respectively).  The curvatures $R_{\alpha_1\cdots \alpha_S\beta_1\cdots \beta_S}$
invariant under (\ref{gauge}) contain $S$ derivatives of the fields \cite{de Wit:1979pe} and are obtained from
$\partial_{\alpha_1\cdots \alpha_S} h_{\beta_1\cdots \beta_S}$ by symmetrizing according to the Young tableau with $S$ columns
and 2 rows\footnote{For more information, see \cite{Fronsdal:1978rb}. The gauge parameters are subject to tracelessness conditions for $S \geq 3$ and the fields are subject to double tracelessness conditions for $S \geq 4$ \cite{Fronsdal:1978rb}, but as studied in \cite{Francia:2002aa,Francia:2002pt,Bouatta:2004kk,Bekaert:2003az}, it is useful to get rid of the tracelessness conditions.  These will not be considered here.}.

We shall now show that the gauge transformations and the gauge invariant curvatures can be naturally reformulated in terms of a differential operator that fulfills $d^N = 0$ with $N=S+1$.

\section{The $N$-complexes $(\Omega_N(\mathbb{R}^D), d)$}

Let $Y^N_p$ be the Young diagram with $p$ cells defined in the following manner:
write the division of $p$ by $N-1$, i.e. ,  write $p = (N -1)n_p + r_p$ where $n_p$ and $r_p$ are
(the unique) integers with $0 \leq n_p$ and $0 \leq r_p \leq N-2$ ($n_p$ is the quotient whereas
$r_p$ is the remainder).  The Young diagram $Y^N_p$ has $n_p$ rows of $N-1$ cells
and the last row with $r_p$ cells (if $r_p \not= 0$). In standard notations, one has $Y^N_p = ((N - 1)^{n_p} , r_p)$, that is
we fill the rows maximally.  The form of a generic $Y^N_p$ Young diagram is illustrated in Fig. 1.
\begin{figure}[H]
\begin{center}
\includegraphics{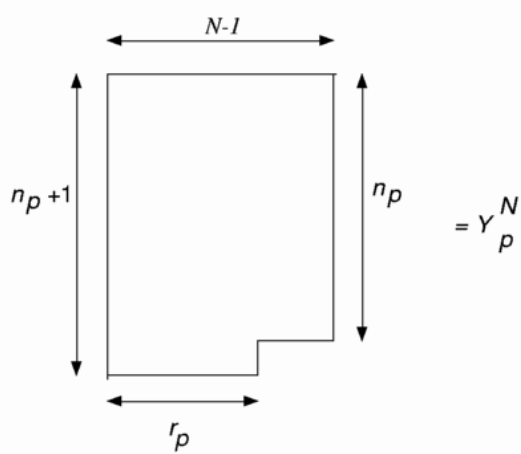}
\caption{A typical Young diagram $Y^N_p$, with $n_p$ rows of $N-1$ cells
and the last row with $r_p$ cells (if $r_p \not= 0$).\label{fig1}}
\end{center}
\end{figure}

Other examples with $N =5$ are given below.\\

\vspace{1cm}
    $Y^5_{3}$ = \begin{tabular}{|c|c|c|}\hline
     & & \\\hline
     \end{tabular} \hspace{1cm}
     $Y^5_{16}$ = \begin{tabular}{|c|c|c|c|}\hline
  &  &  &  \\ \hline
   &  &  &  \\ \hline
   &  &  &  \\ \hline
  &  &  &   \\ \hline
 \end{tabular}
 \hspace{1cm} $Y^5_{17}$ = \begin{tabular}{|c|c|c|c|}\hline
  &  &  &  \\ \hline
   &  &  &  \\ \hline
   &  &  &  \\ \hline
  &  &  &   \\ \hline
      \\
 \cline{1-1}
 \end{tabular}
 \\
 \vspace{1cm}

 $Y^5_{18}$ = \begin{tabular}{|c|c|c|c|}\hline
  &  &  &  \\ \hline
   &  &  &  \\ \hline
   &  &  &  \\ \hline
  &  &  &   \\ \hline
      & \\
 \cline{1-2}
 \end{tabular}
 \hspace{1cm}
 $Y^5_{19}$ = \begin{tabular}{|c|c|c|c|}\hline
  &  &  &  \\ \hline
   &  &  &  \\ \hline
   &  &  &  \\ \hline
  &  &  &   \\ \hline
     & & \\
 \cline{1-3}
 \end{tabular}
 \hspace{1cm}
 $Y^5_{20}$ = \begin{tabular}{|c|c|c|c|}\hline
  &  &  &  \\ \hline
   &  &  &  \\ \hline
   &  &  &  \\ \hline
  &  &  &   \\ \hline
  &  &  &   \\ \hline
 \end{tabular}
 \hspace{1cm}
 \\

\vspace{1cm}

A diagram which is not of $Y^N_p$-type is \\

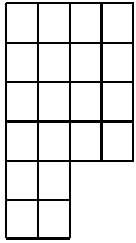
    \begin{center}
    \begin{tabular}{|c|c|c|c|}\hline
  &  &  &  \\ \hline
   &  &  &  \\ \hline
   &  &  &  \\ \hline
  &  &  &   \\ \hline
     & \\
 \cline{1-2}
 & \\
 \cline{1-2}
 \end{tabular}
 \end{center}

\vspace{.2cm}
\noindent since one has put boxes on the 6th row before filling the 5th row.

We call the Young diagrams $Y^N_p$ with $p = (N - 1)n_p$ ``well-filled diagrams". These are rectangular diagrams with $n_p$ rows of $N - 1$ cells each. So, $Y^5_{16}$ and $Y^5_{20}$ above are well-filled, while $Y^5_3$, $Y^5_{17}$, $Y^5_{18}$ and $Y^5_{19}$ are not.

Throughout the following $(x^\mu) = (x^1, \cdots , x^D)$ denotes the canonical coordinates
of $\mathbb{R}^D$ and $\partial_\mu$ are the corresponding partial derivatives which we identify with
the corresponding covariant derivatives associated to the canonical flat linear
connection of $\mathbb{R}^D$. Thus, if $T$ is a covariant tensor field of degree
$p$ on $\mathbb{R}^D$ with components $T_{\mu_1 \cdots \mu_p}(x)$, then $\partial T$ denotes the covariant tensor field
of degree $p + 1$ with components $\partial_{\mu_1}T_{\mu_2 \cdots \mu_{p+1}}(x)$. The operator $\partial$ is a first-order
differential operator which increases by one the tensorial degree.

We denote by $Y^N = (Y^N_p)_{p \in \mathbb{N}}$ the sequence of the Young diagrams $Y^N_p$ ($p \in \mathbb{N}$).  We define $\Omega_N^p(\mathbb{R}^D)$ to be the vector space of smooth covariant tensor fields of degree $p$ on $\mathbb{R}^D$ which have the Young symmetry type $Y^N_p$.  We let $\Omega_N(\mathbb{R}^D)$ be the graded vector space $\oplus_p \Omega_N^p(\mathbb{R}^D)$.  If $N=2$, then $S = N-1 = 1$ and the space $\Omega_2(\mathbb{R}^D)$ is just the space $\Omega(\mathbb{R}^D)$ of differential forms on $\mathbb{R}^D$, i.e., the graded vector space of (covariant) antisymmetric tensor fields on $\mathbb{R}^D$ with graduation induced by the tensorial degree.  The standard exterior differential $d$ on differential forms is the composition of the above $\partial$ with antisymmetrisation, i.e.
\begin{equation} d = \mathbf{A}_{p+1} \circ \partial :
\Omega^p_2(\mathbb{R}^D) \rightarrow
\Omega^{p+1}_2(\mathbb{R}^D) \label{ext} \end{equation}
where $\mathbf{A}_p$ denotes the antisymmetrizer on tensors of degree $p$. One has $d^2 = 0$ because partial derivatives commute. The Poincar\'e  lemma asserts that the cohomology of the complex $(\Omega_2(\mathbb{R}^D), d)$ is trivial, i.e. that one has $H^p(\Omega_2(\mathbb{R}^D)) = \ker(d : \Omega_2^p(\mathbb{R}^D) \rightarrow \Omega_2^{p+1}(\mathbb{R}^D))/d(
\Omega_2^{p-1}(\mathbb{R}^D)) = 0, \forall p \geq 1$, and $H^0(\Omega_2(\mathbb{R}^D)) = \ker(d : \Omega_2^0(\mathbb{R}^D) \rightarrow \Omega_2^{1}(\mathbb{R}^D)) = \mathbb{R}$.  This complex is well-known to be relevant to the description of spin 1 gauge fields.

{}For $N \geq 3$, we generalize the exterior differential by setting $d = \mathbf{Y} \circ \partial$, i.e.,
\begin{equation} d = Y_{p+1} \circ \partial :
\Omega_N^p(\mathbb{R}^D) \rightarrow
\Omega_N^{p+1}(\mathbb{R}^D) \label{deriv} \end{equation}
where $\mathbf{Y}_p$ is now the Young symmetrizer on tensor of degree $p$ associated to the
Young symmetry $Y_p^N$. This $d$ is again a first order differential operator which is
of degree one, (i.e. it increases the tensorial degree by one), but now, $d^2 \not= 0$ in
general. Instead, the following result holds.

\begin{lemma}

One has \begin{equation} d^N = 0 \label{nilpotence} \end{equation}

\end{lemma}

\vspace{.5cm}

\noindent {\bf Proof:}
In fact the indices in one column are antisymmetrized and $d^N T$ involves necessarily
at least two partial derivatives in one of the columns since there are $N$
partial derivatives involved and at most $N - 1$ columns. $\Box$

\vspace{.5cm}

It is clear that $(\Omega_2(\mathbb{R}^D), d)$ is the usual complex of
differential forms on $\mathbb{R}^D$ as was recalled above. The $N$-complex $(\Omega_N(\mathbb{R}^D), d)$ will be simply denoted by $\Omega_N(\mathbb{R}^D)$.

It is easy to write down explicit formulas in terms of components. Consider for instance the case $N = 3$, for which the relevant Young diagrams are those with
two columns, one of length $k$ and the second of length $k-1$ or $k$. A tensor field in $\Omega_3(\mathbb{R}^D)$ is a scalar $T$ in tensor degree 0, a vector $T_\alpha$ in tensor degree 1, a symmetric tensor $T_{\alpha \beta}$ in tensor degree 2. In tensor degree $2k - 1$ ($k \geq 2$), it is described by components $T_{\alpha_1 \cdots \alpha_k \beta_1 \cdots \beta{k-1}}$ with the Young symmetry of the diagram with $k-1$ rows of length 2 and one row of length 1, while in even tensor degree $2k$, it is
described by components $T_{\alpha_1 \cdots \alpha_k \beta_1 \cdots \beta{k}}$ with the Young symmetry of the well-filled rectangular diagram with $k$ rows of length 2. The components of $dT$ are respectively
proportional to $\partial_\alpha T$, $\partial_{(\alpha} T_{\beta)}$, $\partial_{[\alpha_1}T_{\alpha_2]\beta}$ and $T_{\alpha_1 \cdots \alpha_k [\beta_2 \cdots \beta{k}, \beta_1]} + T_{\beta_1 \cdots \beta_k [\alpha_2 \cdots \alpha{k}, \alpha_1]}$
or $\partial_{[\alpha_1}T_{\alpha_2 \cdots \alpha_{k+1}] \beta_1 \cdots \beta{k}}$, where the comma stands for the partial derivative, (. . . ) for
symmetrization and [. . . ] for antisymmetrization. It is obvious that $d^3 = 0$ since
all terms in $d^3T$ involves one antisymmetrization over partial derivatives.  Pictorially, \\

\vspace{.5cm}
    \begin{tabular}{|c|c|}\hline
     &  \\\hline
     \end{tabular} \hspace{.4cm} $ \rightarrow$ \hspace{.4cm}
     \begin{tabular}{|c|c|}\hline
  &   \\ \hline
    $\partial$  \\
 \cline{1-1}
 \end{tabular} \hspace{.4cm} $ \rightarrow$ \hspace{.4cm}
     \begin{tabular}{|c|c|}\hline
  &   \\ \hline
    $\partial$ & $\partial$ \\
 \hline
 \end{tabular} \hspace{.4cm} $ \rightarrow$ \hspace{.4cm}
     \begin{tabular}{|c|c|}\hline
  &   \\ \hline
    $\partial$ & $\partial$ \\
 \hline
 $\partial$  \\
 \cline{1-1}
 \end{tabular} \hspace{.2cm} $=0$

    \vspace{1cm}
and \\

\vspace{.5cm}
    \begin{tabular}{|c|c|}\hline
     &  \\\hline
     \\
 \cline{1-1}
     \end{tabular} \hspace{.4cm} $ \rightarrow$ \hspace{.4cm}
     \begin{tabular}{|c|c|}\hline
  &   \\ \hline
    & $\partial$  \\ \hline
 \end{tabular} \hspace{.4cm} $ \rightarrow$ \hspace{.4cm}
     \begin{tabular}{|c|c|}\hline
  &   \\ \hline
    & $\partial$  \\ \hline
    $\partial$   \\
 \cline{1-1}
 \end{tabular} \hspace{.4cm} $ \rightarrow$ \hspace{.4cm}
     \begin{tabular}{|c|c|}\hline
  &   \\ \hline
    & $\partial$  \\ \hline
    $\partial$ & $\partial$  \\
 \hline
 \end{tabular} \hspace{.2cm} $=0$
    \vspace{1cm}

We can now rewrite the gauge transformations (\ref{gauge}) and curvatures in terms of the generalized $d$.  One has indeed $\delta_\epsilon h = d \epsilon$ and $R = d^S h \equiv d^{N-1} h$.  The gauge invariance of the curvatures of the the spin S gauge field follows from $d^{S+1} \equiv d^N = 0$.  We see therefore that $N$-complexes naturally appear in the description of spin S gauge fields, at least at the linearized level.

\section{Generalized Poincar\'e lemma}

We recall \cite{DV2} that the (generalized) cohomology of the $N$-complex $\Omega_N(\mathbb{R}^D)$ is the family of graded vector spaces $H_{(k)}(\Omega_N(\mathbb{R}^D))$, $k \in \{1, . . . ,N - 1\}$ defined by \begin{equation} H_{(k)}(\Omega_N(\mathbb{R}^D)) = \frac{\ker(d^k)}{\im(d^{N-k})}, \end{equation} i.e., \begin{equation} H_{(k)}(\Omega_N(\mathbb{R}^D)) = \oplus_p H^p_{(k)}(\Omega_N(\mathbb{R}^D))\end{equation} with
\begin{equation}
H^p_{(k)}(\Omega_N(\mathbb{R}^D)) = \frac{\ker(d^k : \Omega^p_N(\mathbb{R}^D) \rightarrow \Omega^{p+k}_N (\mathbb{R}^D))}{d^{N-k}(\Omega^{p+k-N}(\mathbb{R}^D))}. \end{equation}

The following statement generalizes the Poincar\'e  lemma.
\begin{theorem} \label{Poinc}

One has \begin{equation} H^{(N-1)n}_{(k)} (\Omega_N(\mathbb{R}^D)) = 0, \forall n \geq 1 \end{equation} (cohomology trivial at well-filled degrees). {}Furthermore, $H^0_{(k)}(\Omega_N(\mathbb{R}^D))$ is
the space of real polynomial functions on $\mathbb{R}^D$ of degree strictly less than $k$ for $k \in \{1, . . . ,N . 1\}$.

\end{theorem}

This statement reduces to the standard Poincar\'e lemma for $N = 2$ but it is a nontrivial generalization for $N \geq 3$ in the sense that the spaces $H^p_{(k)}(\Omega_N(\mathbb{R}^D))$ are nontrivial for $p \not= (N- 1)n$ and, in fact, are generically infinite dimensional for $D \geq 3$, $p \geq N$.
The second part of the theorem is obvious since the condition $d^kf = 0$ simply states that the derivatives of order $k$ of $f$ all vanish (and there is no quotient
to be taken since $f$ is in degree 0). The proof of the first part of the theorem, which asserts that there is no cohomology for well-filled diagrams, is given in \cite{DuboisViolette:2001jk}.

Useful information on the non-trivial cohomological groups ($p \not= (N-1)n$) can be obtained from a powerful lemma of the general theory of $N$-complexes. This lemma was formulated
in \cite{DV2} in the more general framework of $N$-differential  modules (Lemma 1 of \cite{DV2}) that is of ${\mathbf k}$-modules
equipped with an endomorphism $d$ such that $d^N=0$ where ${\mathbf k}$  is a unital commutative ring. Here we only discuss $N$-complexes of
(real) vector spaces. Let $E$ be a $N$-complex of cochains \cite{DV2} like $\Omega_{N}(\mathbb R^D)$, that is  $E=\oplusinf_{m\in \mathbb N}E^m$ is a graded vector space equipped  with an endomorphism $d$ of degree one such that $d^N=0$ ($N\geq  2$). The inclusions $\ker(d^k)\subset \ker(d^{k+1})$ and  $\im(d^{N-k})\subset \im (d^{N-k-1})$ induce  linear  mappings $[i]:H_{(k)}\rightarrow H_{(k+1)}$ in generalized cohomology
 for $k$ such that $1\leq k\leq N-2$. Similarily the linear mappings  $d:\ker(d^{k+1})\rightarrow \ker(d^k)$ and
 $d:\im(d^{N-k-1})\rightarrow \im(d^{N-k})$ obtained by restriction of  the $N$-differential $d$ induce linear mappings
 $[d]:H_{(k+1)}\rightarrow H_{(k)}$. One has the following lemma (for
 a proof we refer to \cite{DV2}).

\begin{lemma}\label{lemfond}

Let the integers $k$ and $\ell$  be such that $1\leq k$, $1\leq \ell$,
$k+\ell\leq N-1$. Then the
 hexagon of linear mappings
 \[ \begin{diagram} \node{}
\node{H_{(\ell+k)}(E)} \arrow{e,t}{[d]^k} \node{H_{(\ell)}(E)}
\arrow{se,t}{[i]^{N-(\ell+k)}} \node{} \\ \node{H_{(k)}(E)}
\arrow{ne,t}{[i]^\ell}
\node{} \node{} \node{H_{(N-k)}(E)} \arrow{sw,b}{[d]^\ell} \\ \node{}
\node[1]{H_{(N-\ell)}(E)} \arrow{nw,b}{[d]^{N-(\ell+k)}}
\node{H_{(N-(\ell+k))}(E)}
\arrow{w,b}{[i]^k} \node{} \end{diagram} \]
is exact.
\end{lemma}
Since $[i]$ is of degree zero while $[d]$ is of degree one,
these hexagons give long exact sequences.\\

Let us apply the above result to the $N$-complex $\Omega_{N}(\mathbb
R^D)$. For $N=3$, there is only one hexagon as above $(k=\ell=1)$ and,
by using $H^{2n}_{(k)}=0$ for $n\geq 1$, $k=1,2$ it reduces to the
exact sequences
\begin{equation}
 0\stackrel{[d]}{\rightarrow} H ^0_{(1)}\stackrel{[i]}{\rightarrow}
 H^0_{(2)}\stackrel{[d]}{\rightarrow}
 H^1_{(1)}\stackrel{[i]}{\rightarrow} H^1_{(2)}\stackrel{d}{\rightarrow}
 0
 \label{eq.6.6}
 \end{equation}
 and
 \begin{equation}
  0\stackrel{d}{\rightarrow} H^{2n+1}_{(1)}\stackrel{[i]}{\rightarrow}
  H^{2n+1}_{(2)}\stackrel{d}{\rightarrow} 0
  \label{eq.6.7}
  \end{equation}
  for $n\geq 1$. The sequences (\ref{eq.6.7}) give the interesting
  isomorphisms $H^{2n+1}_{(1)}\simeq H^{2n+1}_{(2)}$ while the
  4-terms sequence (\ref{eq.6.6}) allows to compute the finite
  dimension of $H^1_{(2)}$ knowing the one of $H^0_{(1)}$,
  $H^0_{(2)}$ and $H^1_{(1)}$. For $N\geq 3$ one has several hexagons
  and by using $H^{(N-1)n}_{(k)}=0$ for $n\geq 0$, the sequence
  (\ref{eq.6.6}) generalizes as the following $\frac{(N-2)(N-1)}{2}$
  four-terms exact sequences
  \begin{equation}
   0\stackrel{[d]^k}{\longrightarrow}
   H^{k-1}_{(\ell)}\stackrel{[i]^{N-k-\ell}}{\longrightarrow}
   H^{k-1}_{(N-k)}\stackrel{[d]^\ell}{\longrightarrow}
   H^{k+\ell-1}_{(N-k-\ell)}\stackrel{[i]^k}{\longrightarrow }
   H^{k+\ell-1}_{(N-\ell)}\stackrel{[d]^{N-k-\ell}}{\longrightarrow} 0
   \label{eq.6.8}
   \end{equation}
   for $1\leq k,\ell$ and $k+\ell\leq N-1$. There are also two-terms
   exact sequences generalizing (\ref{eq.6.7}) giving similar
   isomorphisms but, for $N>3$, there are other longer exact
   sequences (which are of finite lengths in view of
   $H^{(N-1)n}_{(k)}=0$ for $n\geq 1$). Suppose that the spaces
   $H^m_{(k)}$ are finite-dimensional for $k+m\leq N-1$ and that we
   know their dimensions. Then the exact sequences (\ref{eq.6.8})
   imply that all the $H^m_{(k)}$ for $m\leq N-2$ are
   finite-dimensional and allows to compute their dimensions in terms
   of the dimensions of the $H^m_{(k)}$ for $k+m\leq N-1$. The implications of these relations are further discussed in \cite{DuboisViolette:2001jk}.

\section{Applications}

The generalized Poincar\'e lemma implies interesting results for higher spin gauge fields. The statement $H^S
_{(S)}(\Omega_{S+1}(\mathbb{R}^D)) = 0$ ensures that gauge fields with zero curvatures are pure gauge. This was
directly proved in \cite{Damour:1987vm} for the particular case $S = 3$. The condition $d^{S+1} = 0$ also ensures
that curvatures of gauge potentials satisfy a generalized Bianchi identity of the
form $dR = 0$. The generalized Poincar\'e lemma also implies $H^{2S}_{(1)}(\Omega_{S+1}(\mathbb{R}^D)) = 0$,
which means that conversely the Bianchi identity characterizes the elements of $\Omega^{2S}(\mathbb{R}^D)$ which are curvatures of gauge potentials. This claim for $S = 2$ is the main statement of \cite{Gasqui}.

There is also a generalization of Hodge duality for $\Omega_N(\mathbb{R}^D)$, which is obtained by contractions of the columns with the Kroneker tensor $\epsilon^{\mu_1 \cdots \mu_D}$ of $\mathbb{R}^D$ \cite{DuboisViolette:1999rd,DuboisViolette:2001jk}. When combined with Theorem {\bf \ref{Poinc}}, this duality leads to another kind of results. A typical result is the following one. Let $T^{\mu \nu}$ be a symmetric contravariant tensor field of degree 2 on $\mathbb{R}^D$ satisfying $\partial_\mu T^{\mu \nu} = 0$, (like e.g. the stress energy tensor), then there is a contravariant tensor field $R^{\lambda \mu \rho \nu}$ of degree 4 with the symmetry of the Riemann curvature tensor, such that \begin{equation}
T^{\mu \nu} = \partial_\lambda \partial_\rho R^{\lambda \mu \rho \nu}. \label{key}\end{equation}
In order to connect this result with Theorem {\bf \ref{Poinc}}, define $\tau_{\mu_1 \cdots \mu_{D-1} \nu_1 \cdots \nu_{D-1}} = T^{mu \nu} \epsilon_{\mu \mu_1 \cdots \mu_{D-1}} \epsilon_{\nu \nu_1 \cdots \nu_{D-1}}$.  Then one has $\tau \in \Omega^{2(D-1)}_3 (\mathbb{R}^D)$ and conversely, any $\tau \in \Omega^{2(D-1)}_3 (\mathbb{R}^D)$ can be expressed in this form in terms of a symmetric contravariant 2-tensor. It is easy to verify that $d\tau = 0 $ (in $\Omega_3(\mathbb{R}^D))$ is equivalent to $\partial_\mu T^{\mu \nu} = 0$. On the other hand, Theorem {\bf \ref{Poinc}} implies that $H^{2(D-1)}_{(1)} (\Omega_3(\mathbb{R}^D)) = 0$ and therefore
$\partial_\mu T^{\mu \nu} = 0$ implies that there is a $\rho \in \Omega^{2(D-2)}_3 (\mathbb{R}^D)$ such that $\tau = d^2 \rho$. The latter
is equivalent to (\ref{key}) with $R^{\mu_1 \mu_2 \nu_1 \nu_2}$ proportional to $\epsilon^{\mu_1 \cdots \mu_{D}} \epsilon^{\nu_1 \cdots \nu_{D}} \rho_{\mu_3 \cdots \mu_{D}\nu_3 \cdots \nu_{D}}$ and one verifies that the $R$ so defined has the correct symmetry. That symmetric
tensor fields identically fulfilling $\partial_\mu T^{\mu \nu} = 0$ can be rewritten as in Eq. (\ref{key}) has
been used in \cite{Wald:1986bj,Boulanger:2000rq} in the investigation of the consistent deformations of the free spin two gauge field action.
How duality is implemented at the dynamical level was investigated (at the free level) in \cite{Hull:2001iu,Boulanger:2003vs}.

\section{Conclusions}
We have reviewed how $N$-complexes provide a useful framework for investigating higher spin gauge theories. This was done at the linearized level.  It remains a challenge to fruitfully extend these concepts to the interacting case.

\section*{Acknowledgments}

Work supported in part by IISN-Belgium
(conventions 4.4511.06 and 4.4514.08), by the Belgian National Lottery, by the European
Commission FP6 RTN programme MRTN-CT-2004-005104, and by the Belgian Federal Science Policy Office through the
Interuniversity Attraction Pole P6/11.

\end{document}